# A Review on End-To-End Methods for Brain Tumor Segmentation and Overall Survival Prediction

Snehal R. Rajput[1], Mehul S. Raval[2]

Pandit Deendayal Petroleum University, Gandhinagar, Gujarat, India[1,2]

snehal.rphd19@sot.pdpu.ac.in[1], mehul.raval@sot.pdpu.ac.in[2]

**Abstract**

Brain tumor segmentation intends to delineate tumor tissues from healthy brain tissues. The tumor tissues include necrosis, peritumoral edema, and active tumor. In contrast, healthy brain tissues include white matter, gray matter, and cerebrospinal fluid. The MRI based brain tumor segmentation research is gaining popularity as; 1. It does not irradiate ionized radiation like X-ray or computed tomography imaging. 2. It produces detailed pictures of internal body structures. The MRI scans are input to deep learning-based approaches which are useful for automatic brain tumor segmentation. The features from segments are fed to the classifier which predict the overall survival of the patient. The motive of this paper is to give an extensive overview of state-of-the-art jointly covering brain tumor segmentation and overall survival prediction.

## 1. Introduction

A brain tumor is an accumulation of abnormal cells in the brain. Usually, the cells of the body are replaced by new cells after the period. However, tumor cells do not die and continue to add more cancerous cells to accumulated cancerous tissue. Nowadays, mainly all medical organizations follow the World Health Organization(WHO) classification standards to recognize types of brain tumors. The WHO classifies them based on the origin of cancerous cells or through the behavior of cancerous cells. It categorizes cancerous cells to different grades ranging from Grade I to Grade IV, based on the rate of growth of cancerous cells. Grade I is the least malignant tumor, and Grade IV is the most malignant tumor.

### 1.1 Classification of brain tumor

The brain tumor is classified based on either origin of the cancerous cells or based on the cell behavior. The following sub sections covers their classification.

#### 1.1.1 Brain tumors based on origin of the cancerous cells

In primary brain tumor, the genesis of cancerous cells lies in the brain, and it does not escalate to different body parts. Secondary brain tumors grow in the different parts of the body, and cancerous cells migrate to the brain. Lung, kidney, breast, skin, and colon are the most common organs from which the cancerous cells can spread to the brain. They are also known as metastatic brain tumors.

### 1.1.2   Tumors based on cell behavior

Benign brain tumors are the least invasive type of brain tumor, and they are noncancerous cells. Typically, benign tumors grow at a low pace and do not invade other neighboring tissues. Malignant brain tumors range from noninvasive type to most invasive type. Based on the growth rates, different cancerous cells can invade proximate healthy brain tissue. The most common type of primary brain tumor found in adults is Glioblastomas, also called Glioblastoma multiforme (GBM). Glioblastomas are a fast-growing type that grows from glial cells. High-grade gliomas (HGG) have a higher proliferation rate than low-grade gliomas(LGG) and hence they need intense clinical treatment plans.

Most patients with GBMs die in less than a year, and virtually none has long-term survival chances. These tumors have drawn enormous attention from the research community; for early detection and therapeutic planning of the patient. It can improve the survival tenure of the patient. Also, due to high variability in appearance, shape, and locations in the brain, segmentation in multimodal MRI scans is one of the most critical tasks in biomedical imaging areas. The MRI gives detailed anatomical information of the brain in all three planes; axial, sagittal, and coronal (cf. Fig. 1). It is useful in diagnosing a tumor, treatment plans, aid surgery, and after therapy planning.

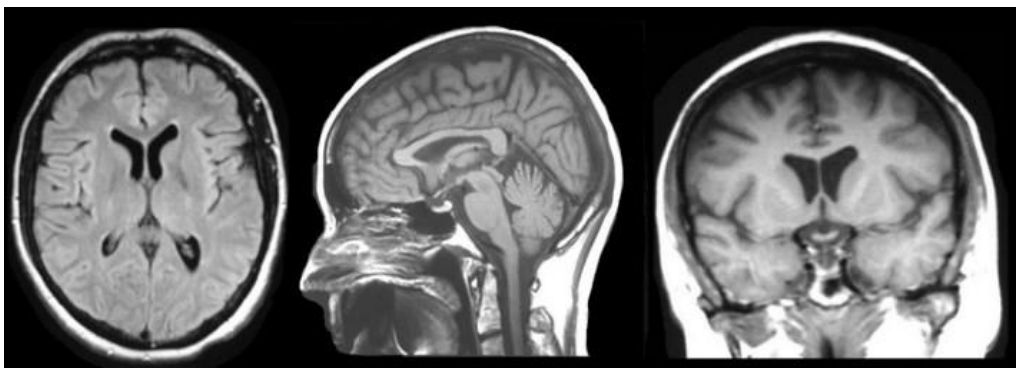

Fig. 1 Brain anatomy in axial, sagittal and coronal planes [1].

## 1.2 Types of MRI sequences

Different types of MRI sequences are; 1. T1 weighted (T1); 2. T2 weighted MRI (T2); 3. Gd-enhanced T1-weighted (T1Gad); 4. Proton Density-weighted MRI (PDW); 5. Fluid Attenuated-Inversion Recovery (FLAIR); 6. functional-MRI; 7. Diffusion-weighted Imaging (DWI). Pulses with short repetition-time (RT) and time-to-echo (TTE) produce T1 sequences. The RT is the time duration between successive pulse sequences applied to the same image slice. The TTE is the time duration between sending the radio frequency pulse and receiving the echo signal. T1-weighted sequence provides a proper differentiation between white matter (WM) and gray matter(GM) while cerebro spinal fluid (CSF) appears black due to lack of signal.

Conversely, by increasing both RT and TTE time, T2 sequences take place. It gives a good contrast between CSF and brain matter, where CSF appears brighter and brain matter appears darker. Typically, T1-weighted and T2-weighted sequences can be ascertained easily by considering CSF as it looks bright on T2 weighted sequence and dark on T1-weighted sequence. Another widely used MRI image scan is FLAIR. Brain image in both T2 weighted images and FLAIR looks similar, with the only difference between the TTE and RT times are quite long in the FLAIR sequence. Like the T2, in FLAIR, gray-matter looks brighter than the white-matter, but CSF looks dark here compared to T2. FLAIR images are susceptible to pathological conditions and make the distinction easy between the CSF and anomalies.

Another commonly used MRI sequence is T1 Gd, obtained by injecting a nontoxic contrast-enhancing agent called Gadolinium. During imaging, when Gadolinium reduces the intensities of T1 images. Hence, it is very bright on T1 weighted sequences. In this paper, we discuss the end-to-end methods used for Brain Tumor Segmentation (BTS) and overall survival (OS) prediction of patients using MRI modality. The schematic diagram of the end-to-end approach is in Fig. 2.

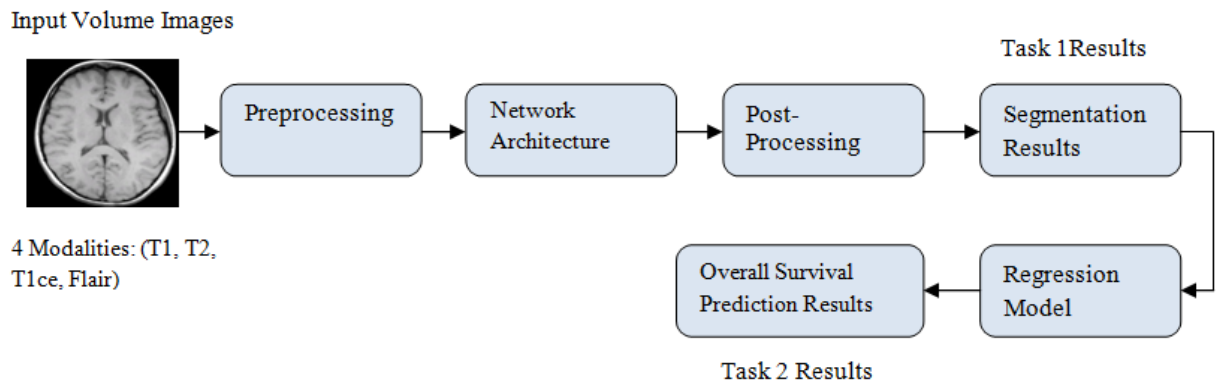

Fig. 2 A schematic diagram for the end-to-end approach for BTS and OS prediction.

Manual diagnosis of tumors from the MRI scans is a time-consuming, complicated, and complex task. Also, the delineation of tumor from healthy tissues relies on the experience of the experts. Hence, there is a need for automation in this task, which can segment tumors with desired precision and accuracy [2]. The rest of the paper proceeds as follows: The second section includes methods for BTS, its challenges, Brain Tumor segmentation (BraTS) challenge tasks, and dataset. The third section has the classification of the Convolutional Neural Network(CNN), with section four covering a literature survey of joint approaches for BTS and OS Prediction. The last section covers the conclusion and future work.

## 2    Brain-Tumor Segmentation Methods

Methods for BTS can be classified based on the level of user intervention i.e., manual-segmentation, semi-automated segmentation, and fully automated-segmentation.

### 2.1 Manual segmentation methods

Manual-Segmentation methods involve manually marking boundaries between healthy and tumor tissues, assigning labels to the region based on anatomic structures. Additionally, it needs broad and in-depth knowledge of anatomy and software tools. Typically, manual segmentation works on contrast-enhanced MRI images. It demarcates tumor areas slice by slice, which restricts expert raters view and hence not able to produce excellent outlined images [3]. Apart from this, manual segmentation outcomes are subject to considerable variability compared to ground truth results. It is due to changes in shape, location, and appearance of cancerous tissues [4]. Manual segmentation will not be optimal due to use of a single image modality. In contrast, the integration of information from multi-modalities can give optimal segmentation results.

### 2.2 Semi automated segmentation methods

In semiautomatic segmentation methods, the intervention of an expert is needed mainly for the initializing of parameters, feeding response back to the model to improve results, and its evaluation [2]. The process can include initializing parameters needed as input for computation such as of the tissue source in the form of seed point selection [5], of the region area [6]. In contrast, feedback response to the model includes recalibrating the input parameters based on the result obtained. Although semi-automated segmentation methods are faster and give competent performance than manual segmentation, improper initialization of parameters may lead to suboptimal segmentation results [3].

### 2.3 Fully automated segmentation methods

In fully automatic methods, the computer segment tumors based on algorithms without any human intervention. Many non-deep learning-based approaches were proposed earlier for tumor segmentation (before 2014) such as random forest classification, extremely random forest-based Markov Random Field (MRF) classification. However, with the advent of deep learning and faster computation, deep learning-based approaches are used for automated segmentation problems.

**2.3.1 Challenges for automatic brain tumor segmentation**

The automatic segmentation of brain tumors is a difficult task due to heterogeneity in tumor tissues. It also suffers from class imbalance, e.g., tumor occupies minimal volume compare to other parts of the brain. This is evident from the Table 1 which shows normal tissues occupying 98% of the brain volume. It also lacks large size annotated multimodal datasets and there is a lack of uniformity among various datasets. The next subsection covers one such important dataset.

Table 1. Class imbalances in the dataset [7].

| Labels | % of brain volume |
|---|---|
| Normal Tissues | 98.00 |
| Necrotic | 00.18 |
| Peritumoral-Edema | 01.10 |
| Non-Enhancing Tumor | 00.12 |
| Active Tumor | 00.18 |

**2.3.2 BraTS-2019 tasks and dataset**

Due to nonuniformity and variations in the dataset, objective evaluation of brain tumor segmentation methods and overall survival prediction are challenging tasks. However, with the advent of standardized benchmark, the BraTS, have become a widely accepted platform for the comparison of various segmentation methods using a common dataset. BraTS 2018 onwards challenge includes three tasks: 1) tumor segmentation, 2) overall survival prediction, and 3) uncertainty estimation for the predicted tumor sub-regions.

The BraTS-2019 [8-10] training dataset includes 335 cases, which contains 259 High-Grade-Glioma and 76 Low-Grade-Glioma cases. The validation dataset consists of 125 cases, along with the ground truth segmentations of each case. For each subject, there are four MRI preoperative scans (T1 weighted, T1 with Gadolinium, T2 weighted, and FLAIR). The ground truth results with annotated labels include Necrotic and Non-Enhancing tumor core NCR/NET ( label-1), edema ( label-2), active tumor (label-4), and 0 for everything else. The dataset pre-processing includes bias-field correction and registration. The scans are skull-stripped and resampled to an isotropic resolution of 1x1x1. Width, height, and depth of each sample are 240, 240, and 155, respectively.

For overall survival prediction, the dataset contains a .csv file, which includes 260 samples for the training set and 126 samples for the validation dataset. Each sample includes age (range: between 27 to 80 years), survival days (range: between 23 to 1592 days), and resection status.

### 2.3.3 Task 1 Similarity measures

For objectifying segmentation tasks, following performance metrics are used:- Dice Similarity Coefficient[8] (Dice-score), Hausdorff distance (since BraTS 2017), specificity, and sensitivity for three central tumor regions; whole-tumor (WT), tumor-core (TC) and enhancing-tumor (ET).

Dice-score: It is defined as follows

$$Dice\ Score\ (S,T) = \frac{2*|S \cap T|}{|S|+|T|}$$

It has a value between 0 and 1. The 0 indicates no match, and 1 signifies the perfect match between predicted and ground truth labels.

Sensitivity (True-Positive-Rate): It is defined as follows:

$$Sensitivity = \frac{True\ Positive}{True\ Positive + False\ Negative}$$

Specificity (True-Negative-Rate): It defined as follows:

$$Specificity = \frac{True\ Negative}{True\ Negative + False\ Positive}$$

The Dice score, sensitivity, and specificity cover the spatial overlapping of the segmented regions and ground truth regions.

A different class of metric calculates the maximum overall surface distance between two given finite element sets $P = \{p_1, p_2, p_3, p_4 \ldots p_k\}$ and $Q = \{q_1, q_2, q_3, q_4 \ldots q_k\}$ known as Hausdorff distance [11].

$$Hausdorff(P,Q) = \max\{h_d(P,Q), h_d(Q,P)\}$$
$$h_d(P,Q) = \max_{p} \in P\left\{\min_{q} \in Q\{d(p,q)\}\right\}$$
$$h_d(Q,P) = \max_{p} \in P\left\{\min_{q} \in Q\{d(q,p)\}\right\}$$

where, $d(p,q)$ is the Euclidean distance between point $p$ and $q$. The functions $h_d(P,Q)$ and $h_d(Q,P)$ are known as Hausdorff distances measured from $P$ to $Q$ and $Q$ to $P$, respectively. The function $h_d(P,Q)$ determine the nearest point in $Q$ for each point in

*P*. The maximum of the values is known as the most mismatched point of *P*. The $Hausdorff\ distance(P, Q)$ is the maximum distances of $h_d(P, Q)$ from *P* to *Q* and from *P* to *Q* for $h_d(Q, P)$.

### 2.3.4 Task 2: Overall survival prediction

Upon segmentation, imaging features are given to machine learning algorithms to predict patient overall survival days. Typically, the prediction task uses imaging and clinical features. Various handcrafted features are; 1. tractographic; 2. spatial like the location of the tumor, the centroid of tumors; 3. first order and second-order statistics; 4. the length of the major axis, minor axis, surface area, and volumetric feature; 5. Geometric features such as enhancing tumor inhomogeneity, tumor surface area irregularity. The geometric features are prudent for predicting overall survival days [12]. Some approaches use Visually Accessible-Rembrandt Images(VASARI); features defined by The Cancer Imaging Archive (TCIA). It is a part of the REMBRANDT project, which had 25 features describing the morphological structure of the brain.

The survival days of patients in the BraTS challenge are classified into three groups of Long-Term-Survivors (≥ 15 Months), Mid-Term-Survivors(≥ 10 Months, and < 15 Months), and Short Term-Survivors (< 10 Months).

### 2.3.5 Task 3: Uncertainty estimation of segmentation

It measures the confidence of the label assigned to each voxel, and the value ranges between 0 and 100, where '100' represents the least confident prediction, and '0' represents the most confident prediction. The voxels exceeding specific predetermined threshold values (T) can be filtered out. The performance of the network architecture is on the resulting voxels Dice score. Removing uncertain voxels or pixels will ideally increase the dice score of the segmentation results. The confidence measures give vital information about the reliability of segmentation results and help to find a critical situation where a medical review is necessary. The confident measure of the voxels can help to identify pixels or voxels that have not segmented, and therefore can be used to ratify the segmentation results [8].

In this paper, we discuss various deep learning-based end-to-end approaches encompassing Task 1 (Tumor segmentation) and Task 2 (Overall survival prediction). The paper also discusses a brief overview of techniques used to estimate the uncertainty of segmentation (Task 3) in Section 5.

## 3  Convolution Neural Networks for segmentation

In recent years, deep learning methods have been used to solve a plethora of problems in different areas of research - most prominently in computer vision, pattern recognition, natural language processing. In many problem areas, deep learning-based approaches have outperformed the previous state of art methods. This achievement is due to the CNNs that can learn useful features from input data, without relying on self-engineered features [13]. For brain tumor segmentation, all the CNNs have contributed remarkably to achieve excellent performance. Apart from all the merits, it has demerits, too, i.e., it lacks interpretability and does not apprehend model uncertainty well [14]. Hence it is critical to design a robust network architecture for segmentation problems. Various deep learning-based CNN architectures are useful for BTS. The CNN architecture is classified based on the input data dimensions and structure of networks.

### 3.1 Classification of CNNs architecture

### 3.1.1 Classification based on structure

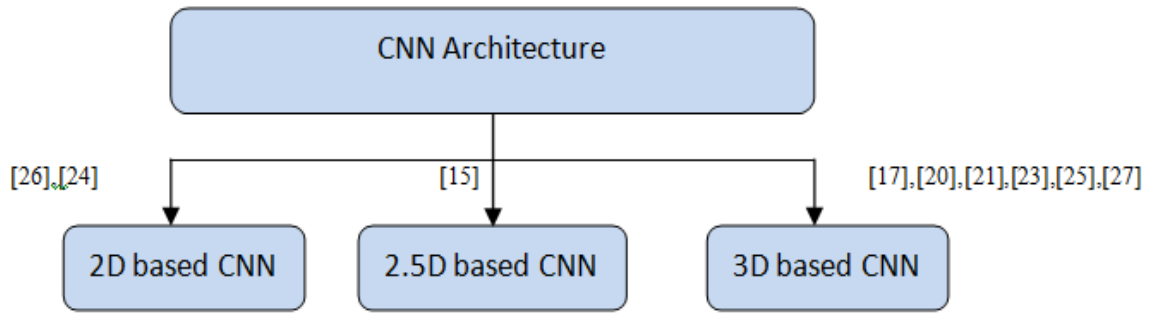

Fig. 3 Classification based on CNN structure.

Various approaches for segmentation use 2D CNN architecture. The slices of 2D images are used as input to perform segmentation tasks or 3D based CNN architecture, use patches of voxels as input to perform segmentation tasks. 2D CNN architecture requires low memory, but it lacks depth information, which will restrict the performance of segmentation [15]. 3D CNN architecture also exploits depth but requires an enormous RAM size to process. It may bound the size of the input patch, channel size, or numbers of features map of the architecture [15]. A 2.5 based CNN approach uses inter-slice images along with intra-slice images. The images from orthogonal views are the input in the network. It is a tradeoff between 2D based CNN and 3D based CNN. 2.5 based CNN can have some advantages over 2D, and 3D based CNN. It can capture inter-slice features, which 2D CNN cannot, and require less memory compared to 3D CNN.

Table 2. Comparison between architectures classified based on input dimensions.

| 2D based CNN architecture | 2.5D based CNN architecture | 3D based CNN architecture |
|---|---|---|
| Merits: requires less memory. | Merits:<br>- can capture depth information through inter-slice images.<br>- requires less memory than 3D CNN.<br>- The performance will be | Merits:<br>- captures depth information.<br>- Accuracy is better than 2D and 2.5D Architecture. |

| | better than 2D CNN architecture. | |
|---|---|---|
| Demerits:<br>- lacks depth information.<br>- Performance may suffer. | Demerits:<br>- It requires more memory than 2D architecture.<br>- Performance is not better than 3D architecture. | Demerits:<br>- requires a considerable size of memory.<br>- more processing is required.<br>- Performance may suffer if the size of memory is limited. |

### 3.1.2 Classification of architecture based on numbers of classifier

The architecture can also be classified based on the number of classifiers; single model or cascaded model or ensemble network model [16]. Each model is based on 2D CNN or 3D CNN.

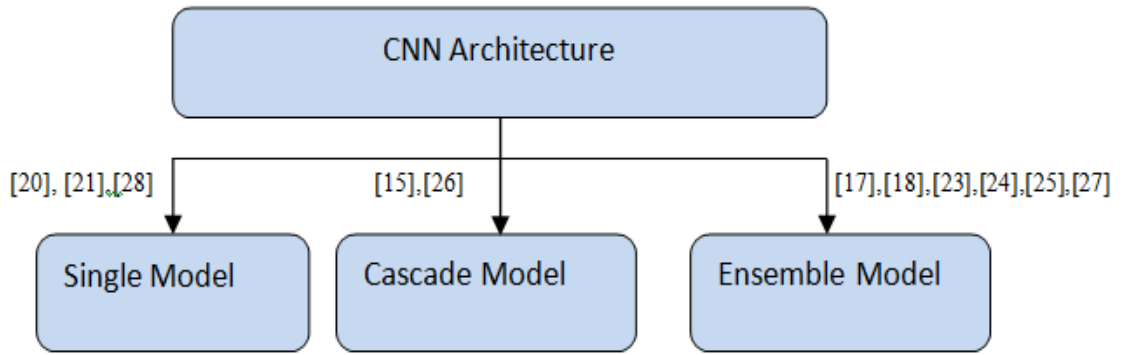

Fig. 4 Classification of CNN architecture based on numbers of classifiers.

Unique model-based approaches include a single network to segment all the three ROI (Regions of Interest), i.e., whole tumor, tumor core, and active tumor. Whereas, in the cascaded model, the segmentation of three regions of interest occurs on three different models. These three networks can be identical or with minor differences in the architecture. In cascade models, the outcome of the first network will be input to the second network. Likewise, the second model's output will be input to its successive model. The approach can successfully control the number of false positives pixels due to the cascade structure. Each successive network will have a restrictive region as input defined by the predecessor network [15]. Ensemble-based models are the collection of various networks with identical architecture. The intuition for the ensemble model is that it provides a more robust result compared to single-model methods. Also, some network architectures may compensate for the limitation of other networks and can help to increase the overall network performance [17]. The next section covers the state-of-the-art covering end to end methods with BTS and OS prediction.

## 4. Joint methods for BTS and OS prediction

The authors [18] proposed an ensemble of U-Net and FCNN for the brain-tumor segmentation task. MRI scans passed as input to both the networks individually, train the network, and the final result is obtained by fusing the result from both the networks. One hundred sixty-three sample scans train the network model, and 66 sample scans validate the network model. The U-NET comprises symmetrical contracting and expansion path, with a skip connection at each level between contracting and expansion path. U-NET captures sufficient contextual information, but it loses local information during down sampling, which is extremely necessary during the expansion path for image construction. Through skip connection at different levels, structural information passes to the respective levels at the expansion path, which is necessary to achieve robust performance[19]. The second network comprises of encoder-decoder architecture, with VGG-11 forms a pre-trained model. The encoding phase includes convolution and max-pooling operations. In contrast, the decoding phase includes the deconvolution operation to obtain the same output size as the input. Unlike the U-NET model, the outcome of each layer forms a feature to the corresponding deconvolutional layer. Finally, the layer with maximum probability gives the segmentation result. For segmentation, cross-validated accuracy was 76.07%, and a mean squared error was 438.54 for the training data. Additionally, validation dataset accuracy was 57.1%, with a mean squared error of 382.96.

For survival prediction, approximately thirty-one thousand features from the three ROI based on texture, histograms, grey-tone difference matrix, co-occurrence matrix, and 3D features are available. Additionally, a novel 10 layers 3D CNN architecture enhances survival prediction accuracy. The new features from MRI inputs and segmentation results combine with the features from the dense layer for the overall survival estimation task. Accuracy on validation and test dataset was 67% and 57% respectively for survival prediction.

Albiol et al. [17] used an ensemble of VGG, two inception models, and fully connected networks. Z-score normalization and data augmentation are preprocessing techniques. The VGG-like model contains four blocks, where each block comprises batch normalization, two 3D Conv layers, and ReLU activation function. In the last layer, softmax performs segmentation. The dense-like model was similar to the VGG model; it includes 20 3D convolution layers, two fully connected layers, and a logit activation function for segmentation. The other two models, inception-2 and inception-3, were like Googlenet with few modifications. It included three blocks of convolution layers with the number of feature parameters. Next, it has inception layers, two fully connected dense layers, and logit function in the last layer for segmentation. The only difference between inception-2 and inception-3 models is the number of inception layers. Each model trains individually, with an 80:20 train to test ratio. Networks were trained using Adam optimizer and with a learning rate = 0.0001 for 40000 iterations. The ensemble model performed better than individual models. On the validation dataset, dice-scores for whole-tumor, tumor-core, and enhancing-tumor were 0.872, 0.760, and 0.751, respectively.

The Ujjwal et al. [20] proposed a single model-based approach that has a three-layered, 3D U-NET model. Each layer includes two Conv operations with kernel size 3x3x3, RELU operation, and batch normalization. For the training model, the 3D patch of size 64x64x64 covers the sample scans. The model trains for 50 iterations and produces four probability maps for each ROI and background. Labels are derived using a maximum probability for each map. Preprocessing of the input data includes bias-field corrections on each image to correct the grayscale heterogeneity of each channel using the N4ITK tool. Also, each modality of scans was normalized to have mean 0 and variance 1.

The post processing includes 3D connected-component analysis which removes false positives in some of the segmentation results. Subsequently, to overcome over-segmentation found in some cases, binary brain masks were produced from brain volume, and with logical AND operation on the segmentation result. It helped to improve the dice score of the segmentation significantly. On the validation dataset, Dice-scores were 0.88, 0.83, and 0.75 for the whole tumor, tumor core, and enhancing tumor, respectively. Overall-Survival-Prediction uses 468 features. It includes first-order statistics, shape features, Gray Level Co-occurrence Matrix (GLCM), and Gray Level Run Length Matrix (GLRLM) features. They are derived using a Pyradiomic package and Cancer Imaging Phenomics Toolkit (CaPTk). The multi-layer perceptron and random forest perform classification and regression. On the validation dataset, the proposed model achieved 57.1% accuracy.

Isensee et al. [21] also proposed a single 3D CNN based U-NET approach in BraTS 2017 and BraTS 2018 challenge. With only a few modifications to the original U-NET approach [19]. Its encoder had five layers, with each layer having two Conv operations, instance normalization, and Leaky RELU as the activation function. The critical difference between batch and instance normalization is that the latter applies it to each instance whereas, former applies it to the whole batch [30]. Several feature maps initialize to 30, double with subsequent layers. For down sampling, max pool operation uses a kernel size of 2x2x2. At the decoder, upsampling uses the same kernel size. Each input image normalizes to zero mean and unit variance of the foreground region. The background region has label 0, whereas the foreground region represents the brain area. A patch size of 128x128x128 captures essential semantic information, and a batch size of two fits GPU memory. A combination of soft Dice and cross-entropy loss trains the network. At the output layer, softmax predicts class labels. Dice scores of the proposed model are 89.51, 70.69, 82.76 for the WT, TC, and ET.

Isensee et al. [22] added a region-based training approach, an additional dataset, and a combination of soft dice and cross-entropy loss. The results of different variations on the baseline model are in [21]. For OS prediction Isensee et al. [22] proposed the Random forest regression (RFR) model [22]. It uses a total of 517 features using the Pyradiomics package. Additionally, they trained an ensemble of 15 multilayer perceptron models (MLP). The average outputs of both the RFR and the MLP ensemble provide final output. The accuracy of the test dataset was 52.6%, with 457.83 RMSE.

McKinley et al. proposed a similar kind of U-NET based ensemble network [23], where densely connected blocks form a basic unit. Instead of the transition layer, a dilated convolution layer increases the receptive field. They introduced a new loss function to train the network, which is a variation of the binary loss function, with an uncertainty factor introduced in it. Further, they introduce focal loss to tackle the impacts of class imbalance. Prior training, input data have been processed and augmented. The dimension of input tensor was: 2x4x5x192x192, which includes batch size, modalities, images from all planes, and spatial dimensions. The performance on the validation dataset with Dice scores is 0.901, 0.854, and 0.795 for the whole tumor, tumor core, and enhancing tumor, respectively.

The [24] proposes the 2D network with the encoder-decoder architecture covering X-Y, Y-Z, and X-Z planes. Instead of max pooling, modified max-pooling preserves the spatial location of max feature values, which in turn helps to maintain smooth boundaries of ROIs at decoder output. All the modalities use a patch size of 128x128. The combined loss function of dice-loss and weighted cross-entropy train all the networks. At the output, all the feature maps fuse to get the result. Invalidation cases, the dice-scores were 0.88 for WT, 0.77 for ET, 0.80 for TC. For survival forecasting, thirty-three semantic features (e.g., necrosis, shape, location) and fifty agnostic features(such as texture, intensity) from the segmentation results are available. These extracted features form input into two layers MLP (multilayer perceptron). The accuracy of the method was 54% on the validation dataset.

Sun L. et al. [25] proposed a combination of both cascade and Ensemble 3D models. The network model includes Wang [15] based cascade model, Isensee [21] based U-NET model, and original U-NET [19] model. The ensemble model gives a competitive performance on the validation dataset with dice-scores of 0.80 for enhancing tumor, 0.90 for the whole tumor, and 0.84 for tumor core. For survival prediction, fourteen features were selected from 4524 features to train a random forest regressor model. These features are first-order statistics, shape, and texture features. In the validation and test case, the accuracy was 46.4% and 61% for survival prediction.

The [26] proposes a cascaded approach were three different 2D U-NETs, one for each region of interest, i.e., for the whole tumor, tumor core, and enhancing tumor. Before training, each image was padded to 256×256 to create ordinary resolution. Each slice is flip using data augmentation to add variations on the dataset images. A patch of size 64x64 was created based on their observation that all tumors are in 64 or fewer adjacent slices. Soft dice loss function trains the network. The summing contours of each model produced the WT contour. The TC contour is available by combining the output of the ET and TC models. The enhancing tumor region was the direct output of the enhancing tumor model. The best result from each model was taken as output after 100 epochs run. The Dice-score for the whole-tumor was 0.87, for tumor core was 0.76, and for enhancing tumor was 0.72.

For overall survival prediction, a three-layer 2D CNN architecture is useful, and the network initializes with orthogonal weights. More recent work has suggested initialization with orthogonal weights or kernels, the generated space tends to display rich features which help to achieve better accuracy. The 2D slices with dimensions

240x240 and ground truth with input are input to this 2D CNN. The resulting representation vector of dimensions 12x12x30, along with patients' age, was passed into an extreme learning machine (ELM) regressor model. ELM is feedforward neural networks, with the parameters of hidden nodes randomly generated (need not be tuned), and by analytically computing output weights. The proposed method accuracy on training data was 0.86 for survival prediction.

Likewise, [27] propose a two-step 3D UNET based cascaded approach. The first network detects the contour map of the tumor, and the second network delineates the tumor detected from the first network into ET, TC, and edema. For all the input images, the brain mask creates differentiating the brain area with nonbrain areas. Non-brain areas are 0. Slices with brain images were normalized. T1, T2, FLAIR, and T1ce-T1 (T1ce image without the T1 image) is the input. The dice-score for validation cases is 0.88 for the whole tumor, 0.71 for enhancing tumor, and 0.75 for tumor core. The survival prediction uses volume-features from all the regions of interest, distance from the brain to the centroid of the tumor, and age is input to the linear regressor model. Using only the "age" feature to train the linear regressor model helped to achieve accuracy of 0.558 on the test dataset and secured the 3rd place in the BraTS challenge 2018.

Myronenko [28] proposed a CNN based encoder-decoder structure, which won the BraTS2018 challenge. The network has a large asymmetrical encoder with five layers to extract in-depth features. The decoder had three layers to reconstruct segmentation. The three-layered variational autoencoder (VAE) regularize distribution. It ensures useful properties to generate a new sample from the distribution during training. Each layer consists of RESNET like blocks. The block has group normalization, RELU, Conv operation, and identity skip connection. The number of blocks in each layer and the number of feature maps increases. Following pre-processing, augmentation is useful. In essence, normalization of all images, images were generated by applying random intensity shift, rescaling factor, and a random axis mirror flip for all the three axes. The group normalization(GN) is better than batch normalization(BN) for small-batch, i.e., for 1 or 2. GN divides the channels into groups and normalizes within the group computing the mean($\mu$) and variance ($\sigma$) [29]. GN's computation is not dependent on the size of the batch, and its accuracy is stable for even larger batch sizes. The architecture follows the typical U-NET architecture of successively reducing image size by two and increasing feature map size by 2. The dimensions of the kernel used in the convolutions layer were 3x3x3, with 32 feature maps initially. The output dimension of the encoder was 256x20x24x16.

The structure of the decoder is similar to the encoder, where each block reduces feature size by two and increases the image dimension by two leads spatial size the same as the original image size. Following it is a 1x1x1 Conv layer with three feature maps for each ROIs and a sigmoid as the activation function. Three dice loss functions together form the loss function. $Loss_{dsice}$ is between decoder segmentation results and ground truth; $L_2$ loss is between a VAE image and the input image. $Loss_{KL}$ is Kullback Leibler divergence between the estimated distribution and a prior distribution. Validation results of 10 ensembled models using data Augmentation were 0.82, 0.91, and 0.86 average dice scores for ET, WT, and TC, respectively. The proposed model achieved the first position in the Brats-2018 Challenge, and on the test cases, dice-scores were 0.76, 0.88, and 0.81 for enhanced tumor core, whole tumor, and tumor core, respectively.

Table 3. Comparison of the end to end approaches for brain tumor segmentation and Overall survival prediction methods (BRATS 2018 & 2019).

| Author | Preprocessing | Segmentation | Post Processing | OS Prediction | Results for Segmentation | | | Results for OS |
|---|---|---|---|---|---|---|---|---|
| | | | | | Mean Dice WT | Mean Dice TC | Mean Dice ET | |
| Shboul et al. [18] | - | The ensemble of UNET and FCN, with VGG16 as pre-trained model | - | Ten layers 3D CNN model | The validation set accuracy: 57.1% | | | Validation dataset: 57% |
| Albiol et al. [17] | Z-score normalization and data augmentation | Ensemble of VGG-like, Dense connected model, Inception-2 and Inception3 model | - | - | Validation set accuracy 0.881 Test accuracy: 0.850 | Validation set accuracy 0.777 Test accuracy: 0.740 | Validation set accuracy 0.773 Test accuracy: 0.723 | - |
| Ujjwal et al. [20] | Bias field correction and z-score normalization - No data Augmentation | 3D UNET | Connected Component Analysis, logical AND operation between brain map and segmentation result | Multilayer perceptron and random forest regression | Validation dataset:0.88 | Validation dataset:0.83 | Validation dataset:0.75 | Validation dataset accuracy: Random forest: 37.5% MLP:57.1% |
| Caver et al. [26] | Z-score normalization and data augmentation | Three 2D UNET for each region of Interest | - | Three layer 2D CNN | 0.878 | 0.76 | 0.724 | Training dataset:86.4% Validation set:60.7% |
| Isensee et al. [21][22] | Z-score normalization and data augmentation: random rotations, random scaling, random elastic deformations, gamma correction | 3D UNET | for some cases of LGG which do not contain enhancing tumor: replacement of enhancing tumor voxels with necrosis if the total the number of | The ensemble of 15 Multilayer perceptron and random forest regression | Validation dataset: 89.51 | Validation dataset: 70.69 | Validation dataset: 82.76 | Test dataset: 52.6% |

| | | | predicted enhancing tumor < T (threshold value.) | | | | | |
|---|---|---|---|---|---|---|---|---|
| **Author** | **Preprocessing** | **Segmentation** | **Post Processing** | **OS Prediction** | **Results for Segmentation** | | | **Results for OS** |
| | | | | | **Mean Dice WT** | **Mean Dice TC** | **Mean Dice ET** | |
| McKinley et al.[23] | Data Augmentation: random flipping, random rotation, random shift and scaling | 2D UNET | - | - | Validation dataset:0.901 | Validation dataset:0.85 | Validation dataset: 0.795 | - |
| Weninger et al. [27] | normalization on brain mask | Cascaded 3D UNET | - | linear Regressor | Validation dataset: 0.889 | Validation dataset: 0.758 | Validation dataset: 0.712 | Test dataset: 55.8% |
| Myronenko A. [28] | Z-score normalization data augmentation: intensity shift, scaling, random axis flip | UNET + Variational Autoencoder | - | - | Validation dataset: 0.9042 | Validation dataset: 0.8596 | Validation dataset: 0.8145 | - |
| Agravat R. et al. [38-39] | z-score normalization, removal of last ten slice | 2D UNET | - | Random Forest Regressor | Validation dataset:0.7 | Validation dataset:0.63 | Validation :0.6 | Validation accuracy: 51.7% |
| Sun L. et al. [25] | Z-score normalization data augmentation: random flip, random Gaussian noise | An ensemble of 3 3D network | - | Random Forest Regressor | Validation dataset: 0.90 | Validation dataset: 0.84 | Validation dataset: 0.80 | Validation accuracy: 46.4% Test accuracy: 61% |
| Banerjee S. et al. [24] | - | The ensemble of 3 2D network | - | MLP | Validation dataset: 0.88 | Validation dataset: 0.80 | Validation dataset: 0.77 | Validation accuracy: 54% |
| Feng et al. [35] | bias correction algorithm, denoising method to reduce noise, normalization | The ensemble of 6 3D UNET | - | Nine features, linear regressor model | 0.90 | 0.83 | 0.79 | Test case accuracy: 32.1% |

## 5. Task 3 Uncertainty estimation of segmentation result

Uncertainty is through a dataset (also called as "Aleatoric") or network structure (also called as "epistemic" uncertainty) [36]. Aleatoric is model to "homoscedastic" or "heteroscedastic" uncertainty. In homoscedastic, the variation of noise is constant on multiple input samples, whereas in heteroscedastic, it differs for different samples [36]. In BTS, since the intensity values of MRI images are not uniformly distributed, we preprocess the dataset to reduce the uncertainty. Aleatoric uncertainty is removed by incorporating more data, whereas epistemic uncertainty can overcome by incorporating more data.

Since deep neural models are not able to apprehend uncertainty well, various Bayesian deep learning methods are useful but which in turn increases complexity and additional computation cost [31]. Alternatively, different methods to approximate the posterior probability over the weights, such as Laplace approximation, MCMC method (Markov chain Monte Carlo), variational inference Bayesian network, and many others as mentioned here [37] is useful. For semantic segmentation, the variational inference is useful to approximate the posterior probability over the weights. Test time dropouts were used to sample from distribution to calculate uncertainty for in-door and out-door scene understanding [34]. Jungo et al. [32] proposed a variational inference approach to estimate uncertainty for brain tumor segmentation tasks. Dropout at test time is useful to sample networks randomly to calculate the average probability of voxel labels given test inputs over the sampled networks. Entropy approximates the obtained probability vectors.

Eaton-Rosen Z. et al. [33] also used a similar approach to calculate uncertainty for segmentation tasks. They also used Dropout at test time to calculate the average probability of labels given the test dataset. The variance approximates the obtained probability vectors. Wang et al. [15] proposed the method to capture Aleatoric uncertainty using test time data augmentation. They perform rotation along all the axis, scaling, and flipping on the volume images to add variability to the dataset (during training and testing). N variations of input image were obtained through sampling to calculate the probability of labels given test images. The majority vote approximates the obtained probability vectors.

## 6. Conclusion and future work

Automatic brain tumor segmentation and overall survival prediction is a critical and challenging task, which can aid experts for better diagnosis, treatment, and surgical planning. This aid can improve the life expectancy of patients. In this paper, we studied various network architectures and found that ensemble models excel in other network architectures for tumor segmentation. With the reported performance of the methods investigated in this paper, we can state that deep learning-based approaches have the potential to fulfill desirable benchmarks for brain tumor segmentation, with the help of more substantial datasets and data augmentation. For the overall survival prediction task, current approaches are not able to get desirable results because of the lack of ample data available for survival prediction. Deep learning-based methods for survival prediction is used to get competitive results, provided with a large dataset. Further improvements are made by;

1. implementing robust network architectures;

2. data augmentation to fine-tune the multiple dataset;

3. including uncertainties information for correct segmentation;

4. adding complementary information from other MRI scans or other imaging scans such as functional MRIs, Diffusion-weighted Imaging (DWI), Positron Emission Tomography (PET). For survival prediction, incorporating more features and samples through clinical collaboration will improve the prediction accuracy significantly.